\documentclass[intlimits,twoside,a4paper]{article}

\usepackage{amsmath,amssymb}
\usepackage{graphicx}

\usepackage[T2A]{fontenc}
\usepackage[cp1251]{inputenc}

\usepackage[eqsecnum]{cmpj3}

\issue{2018}{21}{4}{43705}
\doinumber{10.5488/CMP.21.43705}
\title[Theoretical study of the structural stability]%
{Theoretical study of the structural stability, electronic and magnetic properties of XVSb ($\text{X}=\text{Fe}$, Ni, and Co) half-Heusler compounds%
}

\author[M. Mokhtari \textit{et al.}]{M. Mokhtari\refaddr{label1,label2},
         F. Dahmane\refaddr{label2},  G. Benabdellah\refaddr{label3},  L. Zekri\refaddr{label1}, S. Benalia\refaddr{label2}, N. Zekri\refaddr{label1} }
\addresses{
\addr{label1} Universit\'e des Sciences et de la Technologie d'Oran Mohamed Boudiaf, USTO-MB, LEPM, \\BP 1505, El M' Naouar, 31000 Oran, Algeria
\addr{label2} D\'epartement de SM, Institut des Sciences et des Technologies, Centre Universitaire de Tissemsilt,\\ 38000 Tissemsilt, Alg\'erie 
\addr{label3} Laboratoire de physique Computationnelle des Materiaux, Universit\'e de Sidi Belabes 22000, Alg\'erie
}

\date{Received July 17, 2018, in final form September 30, 2018}

\begin{document}

\maketitle
\begin{abstract}
The structural, electronic and magnetic properties of half-Heusler compounds XVSb (X $=$ Fe, Co and Ni) are investigated by using the density functional theory with generalized gradient approximation (GGA), and Tran-Blaha modified Becke-Johnson (TB-mBJ) exchange potential approximation. It is found that the half-metallic gaps are generally reasonably widened by mBJ as compared to the GGA approximation. The magnetic proprieties of XVSb (X $=$ Fe, Co and Ni) are well defined within mBJ with an exact integer value of magnetic moment. The band gaps given by TB-mBJ are in good agreement with the available theoretical data. The FeVSb exhibits a semiconductor nature. The CoVSb and NiVSb present half-metallic behaviour with total magnetic moment of $1\mu_\text{B}$ and $2\mu_\text{B}$ in good agreement with Slater-Pauling rule. These alloys seem to be a potential candidate of spintronic devices.
\pacs 71.22.+i, 72.25.-b, 75.75.-c

\end{abstract}

\section{Introduction}
A motivating group   of  ternary  alloys  having  chemical formula XYZ named as half-Heusler (HH) compounds, crystallize in the face centered cubic  structure (space group F$\overline{4}$3m; No.~216) \cite{1, 2}, where X is rare-earth or transition metal, Y is transition metal and Z is the main group element \cite{3}. These materials  have  been  of  large  interest  to  both  experimental and theoretical   researchers since they were first studied by Andreas Heusler \cite{4}. 
Spintronics is a promising field in nanoscale electronics which uses  the  spin  of  electrons,  rather  than  an  electric  charge,  to encode  and  process  data \cite{5}.
Half-metallic ferromagnets (HMFs) are characterized by a semiconductor band structure for one-spin direction with a clear energy gap, while the band structure of the other spin direction is metallic  \cite{Mokh, 6}. The electronic density of states (DOS) at $E_\text{F}$ of an ideal  half-metal is composed of only one spin direction resulting in a very high spin polarization ratio at Fermi level. Principally, these compounds (half-metals) are efficient to employ in the conduction of spin polarized current due to its great sensitivity to the applied magnetic field. These types of materials have many technological applications, such as spintronics \cite{7, 9}  and spin injection to semiconductors \cite{10,12}. 

Many research papers focus on the studies and the analysis of physical properties of half-Heusler compounds at an ambient pressure; e.g., structural as well as optoelectronic  properties of CoVSb half-Heusler are studied in \cite{30}. Electronic and magnetic characteristics of different half-Heusler alloys are calculated in \cite{2} while thermodynamics and elastic properties of NiVSb are investigated by Gu et al.~\cite{29}.

In this paper we report on the basic electronic and magnetic properties of XVSb (X $=$ Fe, Ni, and Co) half-Heusler compounds. Therefore, our main objective is to perform an investigation of structural stability, electronic and magnetic properties of XVSb (X $=$ Fe, Ni, and Co) within both generalized gradient approximation (GGA) in the scheme of Perdew-Burke-Ernzerhof (PBE) and Tran-Blaha (TB) modified Becke and Johnson (mBJ) potential \cite{13}  exchange correlation potential of the  density functional theory (DFT). The mBJ exchange potential, has been demonstrated to give a sufficiently exact gaps for wide-band-gap insulators and to ameliorate half-metallic gaps for half-metallic materials \cite{14, 15}.

\section{Method description}

In this work, the first principles calculations are realized   using  the  full potential linearized augmented plane wave (FP-LAPW) method within the framework of the density functional theory (DFT) \cite{16}  as implemented in the WIEN2k code \cite{17}. The convergence of the basis set was controlled by a cut-off parameter $ K_\text{max} = 7/R_\text{MT}$, where $K_\text{max}$ is the largest reciprocal lattice vector used in the plane wave and $R_\text{MT}$ is the smallest of all atomic sphere radii. For the Brillouin zone (BZ) integration, the tetrahedron method with a 47 special $k$ points in the irreducible wedge (1000~$k$-points in the full BZ) was used to construct the charge density in each self-consistency step.
The magnitude of the largest vector in the charge density Fourier expansion ($G_\text{max}$) was 12~(a.u.)$^{-1}$, and we select the charge convergence as $0.0001e$ during self-consistency cycles. The cut-off energy was taken $-6.0$~Ry. This energy defines the separation of the valence and core states. These parameters confirm good convergences for the total energy.

\begin{figure}[!b]
\centerline{%
\includegraphics[width=9.2 cm]{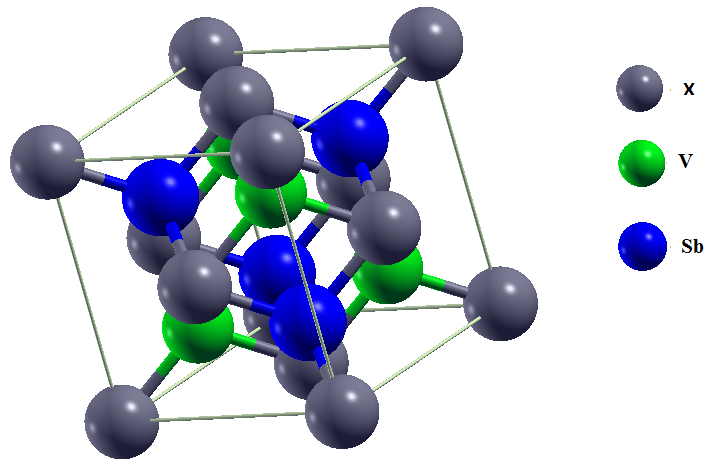}
}
\caption{(Colour online) Crystal structure of half-Heusler XVSb (X $=$ Fe, Co and Ni) with phases $\alpha$.}
\label{fig1}
\end{figure}

The muffin-tin radii (MT) are $2.18$~a.u. for V and $2.24$~a.u for Sb, Fe, Co and Ni. 
To determine the fundamental-state properties, the total energy as a function to the cell volume is fitted to the Murnaghan equation of state \cite{18}, as shown in: 
\begin{equation}
E(V)=E_{0}(V)+\frac{BV}{B'(B'-1)}\left[B\left(1-\frac{V_{0}}{V}\right)+\left(\frac{V_{0}}{V}\right)^{B'}-1\right],
\label{moneq}
\end{equation}
where $E_0$ is the minimum energy at $T = 0$~K, $B$ is the bulk modulus, $B'$ is the bulk modulus derivative, and $V_0$ is the equilibrium volume.
The half-Heusler compounds XYZ crystallizes in the face centered cubic (fcc) (figure~\ref{fig1}) structure with the space group F$\overline{4}$3m (No.~216). Y and Z atoms are located at $4a (0,0,0)$ and $4b (1/2,1/2,1/2)$ position forming the rock salt structure arrangement. The X atoms occupy $4c (1/4,1/4,1/4)$ and $4d (3/4,3/4,3/4)$. When the Z atomic positions are empty, the structure is analogous to the zinc blend structure which is common for a large number of semiconductors.

\section{Results and discussions}

\begin{figure}[!b]
\centerline{%
\includegraphics[width=13 cm]{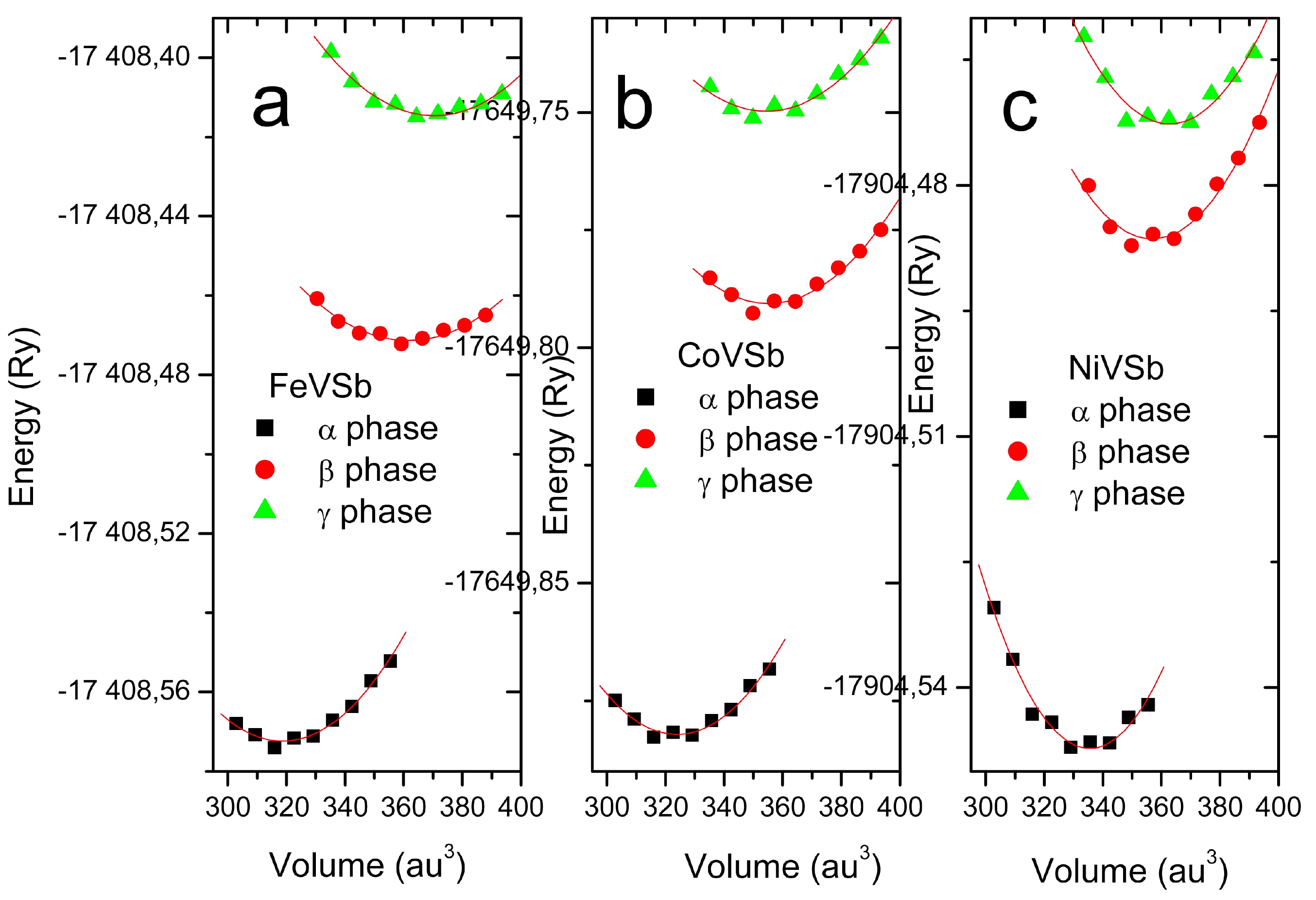}
}
\caption{(Colour online) The total energies per unit cell as a function of the volume for $\alpha$, $\beta$ and $\gamma$ phase  for (a) FeVSb, (b)  CoVSb  and  (c) NiVSb   in the magnetic phase.}
\label{fig2}
\end{figure}

The half-Heusler compounds have a general formula XYZ, the X and Y atoms represent $d$-electron transition metals, and Z denotes a $sp$-electron element. These compounds crystallize in a noncentrosymmetric cubic structure (space group No.~216,   C1b) which is  ternary ordered. Moreover, in the XYZ structure, the X atom has four Y atoms and four Z atoms as the nearest neighbors, whereas the Y and Z atoms have only four X atoms in their nearest neighbor coordination.
 In the half-Heusler C1b structure, the XVSb adopts three possible arrangements named phase $\alpha$, phase $\beta$ and phase $\gamma$ in theory, and these atomic arrangements are possible, which should be distinguished, particularly if electronic structure calculations are executed, since the correct site assignment is important for the obtained electronic structure \cite{19}. In order to define the correct arrangement of the atomic positions in the crystal and the ground state properties, such as the equilibrium lattice constant ($a$), bulk modulus ($B$) and its pressure derivative ($B'$), the structural optimization of three XVSb (X $=$ Fe, Co and Ni) compounds was performed by minimizing the total energy with respect to the lattice parameter variation.  
The total energy curves as a function to the volume for the phases $\alpha$, $\beta$ and $\gamma$ is presented in figure~\ref{fig2} for FeVSb, NiVSb and CoVSb [(a), (b) and (c), respectively)]. From this figure, and for the three compounds, the calculated results show that the phase $\alpha$  disposition is energetically better with a magnetic ground state for all alloys XVSb (X $=$ Fe, Co and Ni) than $\beta$ and $\gamma$ phases because it has the lowest energy. The lattice constant of equilibrium is $5.7976$~{\AA} for FeVSb, $5.8292$~{\AA} for CoVSb and $5.9032$~{\AA} for NiVSb. This is a good agreement with theoretical studies \cite{20, 21, 29}.
The obtained results are given in table~\ref{tab1} and show that the lattice parameter increases with an increasing atomic number of X atom (X $=$ Fe, Co, Ni). As a consequence, the bulk modulus decreases, in accord with the known  link  between  $B$  and  the  lattice  constants: $B\propto  V_{0}^{-1}$, where $V_0$  is the unit cell volume.

For comparison, Ahmad et al. \cite{23}  found that the lattice constant of SbSrX half-Heusler alloys augments in the following sequence:  $a_0 (\text{SbSrC}) < a_0 (\text{SbSrSi}) <a_0 (\text{SbSrGe})$. As Sb and Sr atoms are the same in the three compounds, this result can be easily explained by taking into consideration the atomic radii of C, Si, Ge: the lattice constant increases with an increasing atomic size of the X element in SbSrX compounds. 

 \begin{table}[!t]
\caption{Equilibrium lattice constant $a$~(\AA), the bulk modulus $B$~(GPa) and the minimum energy $E_\text{min}$~(Ry), the gap energy (eV) and the magnetic moment for XVSb (X $=$ Fe, Ni, Co).}
\label{tab1}
\vspace{2ex}
\begin{center}
\begin{tabular}{|c|c||c|c|c|c|c|c||}
\hline\hline
&\phantom{$1^{1\strut}$} &$a$~(\AA)&$B$~(GPa)&$B'$&$E_\text{min}$~(Ry)&$E_g$~(eV)&$m~(\mu_\text{B})$\\
\hline
\hline
\raisebox{-2.9ex}[0pt][0pt]{FeVSb}
      & \raisebox{-1.5ex}[0pt][0pt]{GGA}& 5.7976&  156.1322&  4.9062&  $-$17411.534&  0.41&  $-$0.00029\strut\\
      & & 5.78 \cite{20}&  &  &  &  0.38 \cite{20}&  \strut\\
\cline{2-8}
      & mBJ& &  &  &  &  0.74&  0\strut\\
\hline
\raisebox{-2.9ex}[0pt][0pt]{CoVSb}
      & \raisebox{-1.5ex}[0pt][0pt]{GGA}& 5.8292&  137.9190&  5.6156& $-$17652.871& 0.80& 0.99994\strut\\
& & 5.80 \cite{21}&  &  &  &  0.38&  \strut\\
\cline{2-8}
      & mBJ& &  &  &  &  0.92&  1.00018\strut\\
\hline
\raisebox{-2.9ex}[0pt][0pt]{NiVSb}
      & \raisebox{-1.5ex}[0pt][0pt]{GGA}& 5.9032& 121.4953&  5.5563&  $-$17907.570& 0.36& 2.00289\strut\\
& & 5.786 \cite{29}&  &  5.085 \cite{29}&  & &  \strut\\
\cline{2-8}
      & mBJ& &  &  & &  0.62&  2.00032\strut\\
\hline\hline
\end{tabular}
\end{center}
\end{table}

The energy of formation of the solid is the difference between the energy of a crystal and its constituent as solid phases at zero temperature, which is given by:
\begin{equation}
\Delta H=E(\text{XVSb})_\text{tot}^\text{bulk}-E(\text{V})_\text{tot}^\text{bulk}-E(\text{Sb})_\text{tot}^\text{bulk}-E(\text{X})_\text{tot}^\text{bulk},
\label{3.1}
\end{equation}
where $E(\text{XVSb})_\text{tot}^\text{bulk}$ is the total energy of the alloy. $E(\text{V})_\text{tot}^\text{bulk}$, $E(\text{Sb})_\text{tot}^\text{bulk}$ and $E(\text{X})_\text{tot}^\text{bulk}$ are the energies of the fundamental state per atom of each elemental bulk for V, Sb, Fe, Co and Ni. We find that the formation energies of the stoichiometric XVSb (X $=$ Fe, Co and Ni) in half-Heusler  structure are $-1.23$, $-4.29$ and  $-1.17$~Ry, respectively, which confirms the stability of the alloy in its half-Heusler structure \cite{24, 25}.

\begin{figure}[!t]
\centering
\includegraphics[width=11.5 cm]{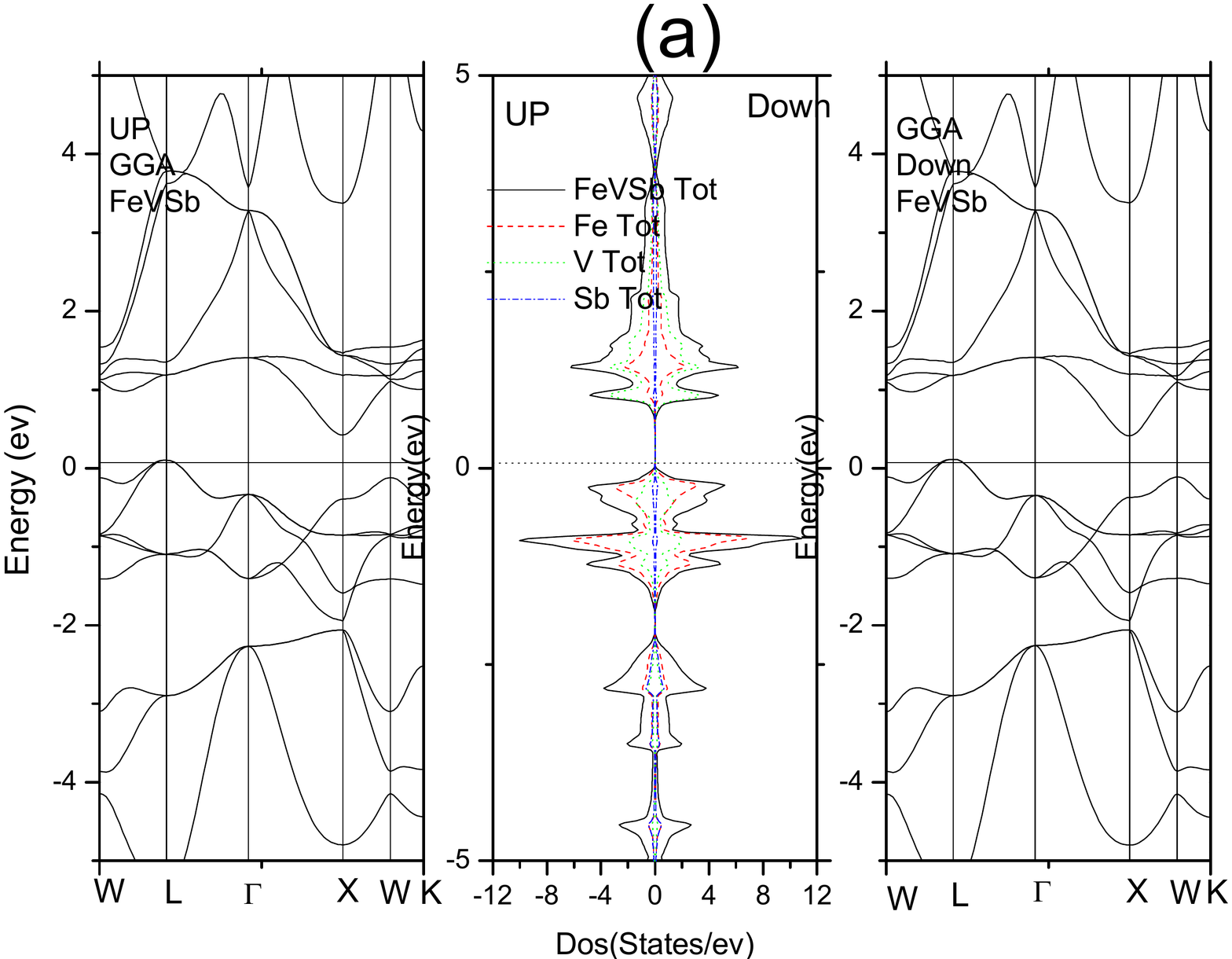}\\
\includegraphics[width=12.3 cm]{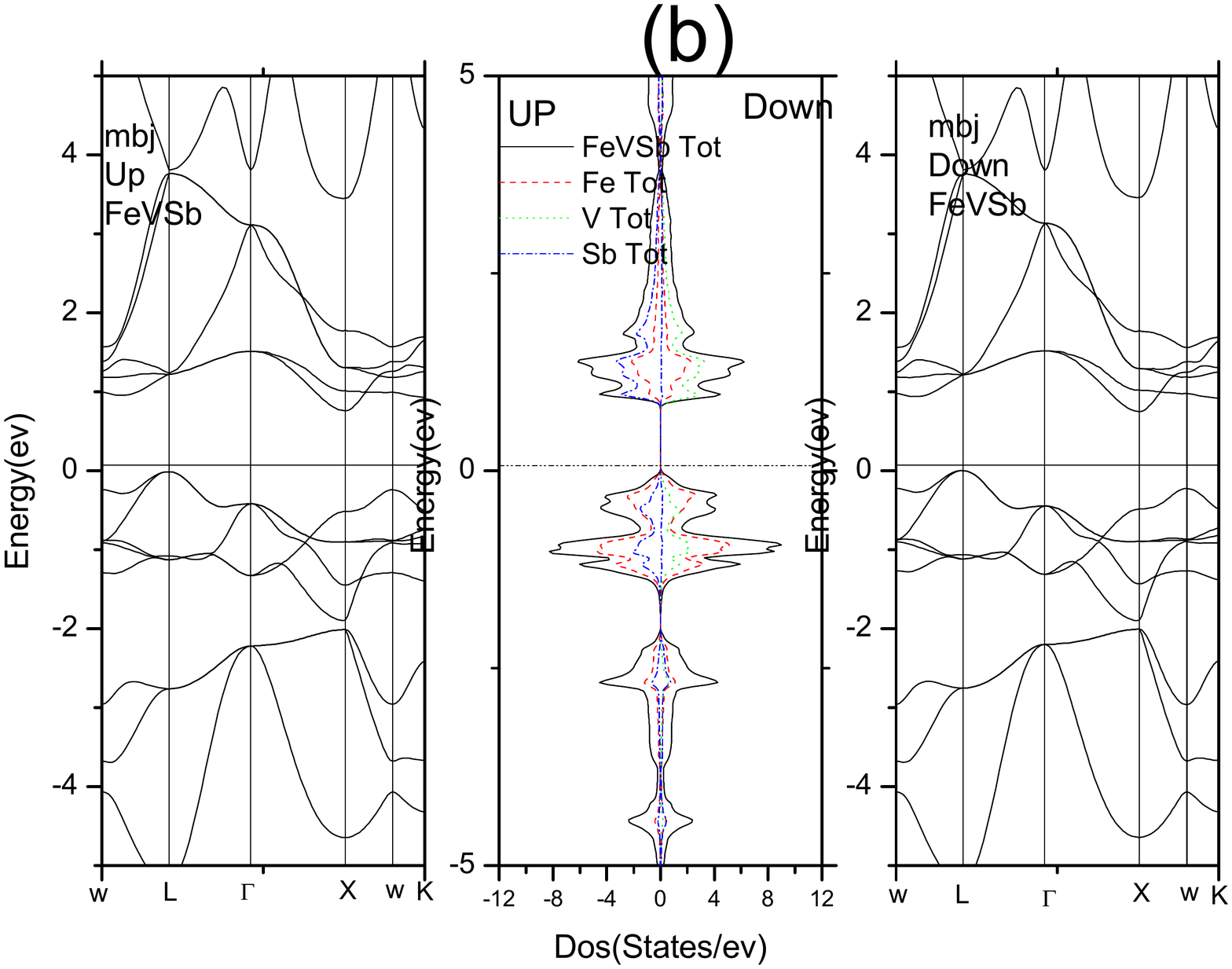}
\caption{(Colour online) The spin-projected total density of states of FeVSb with the corresponding band structures for (a) GGA and (b) TB-mBJ approximations.}
\label{fig3}
\end{figure}

\begin{figure}[!t]
\centering
\includegraphics[width=11.5 cm]{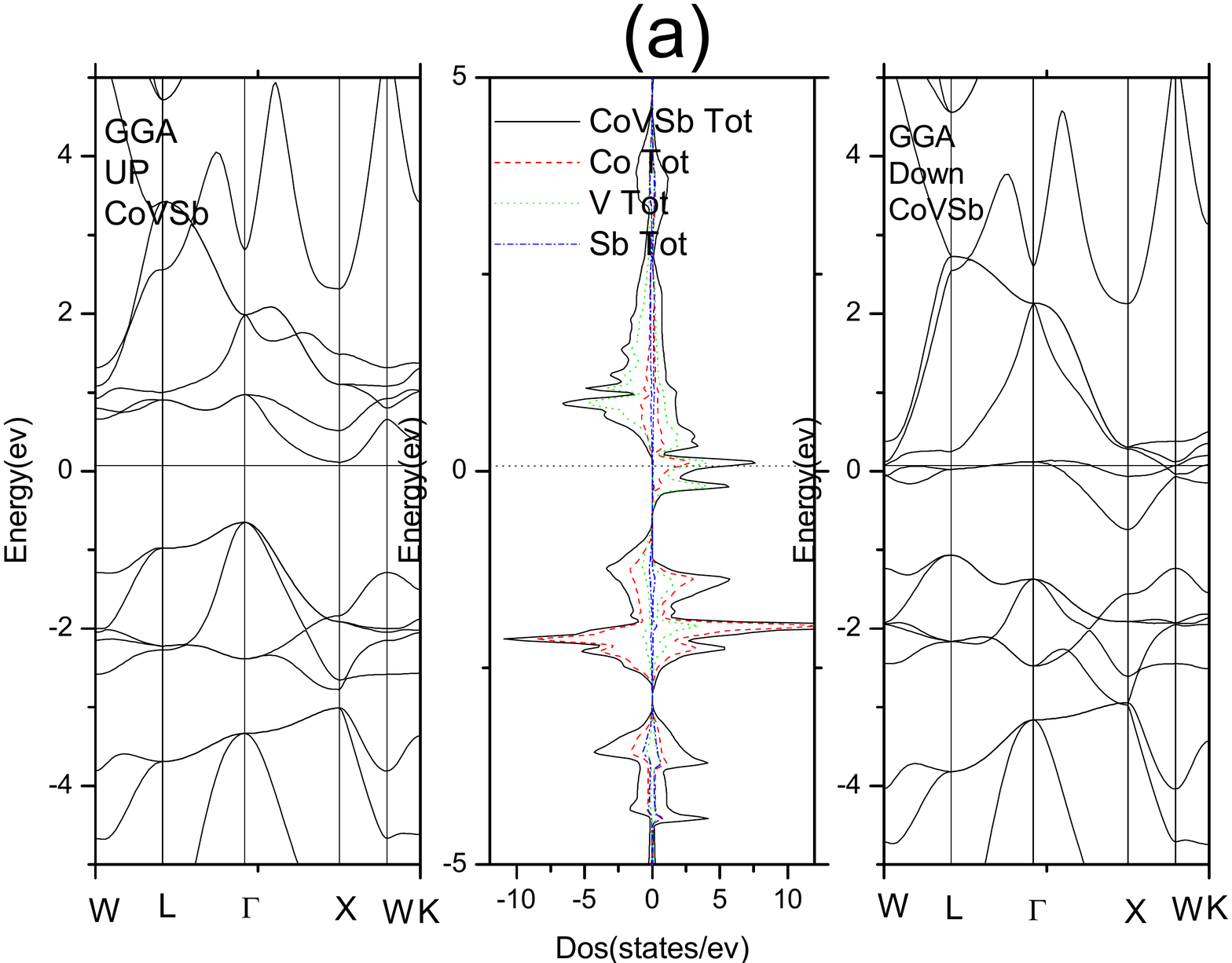}\\
\includegraphics[width=12.3 cm]{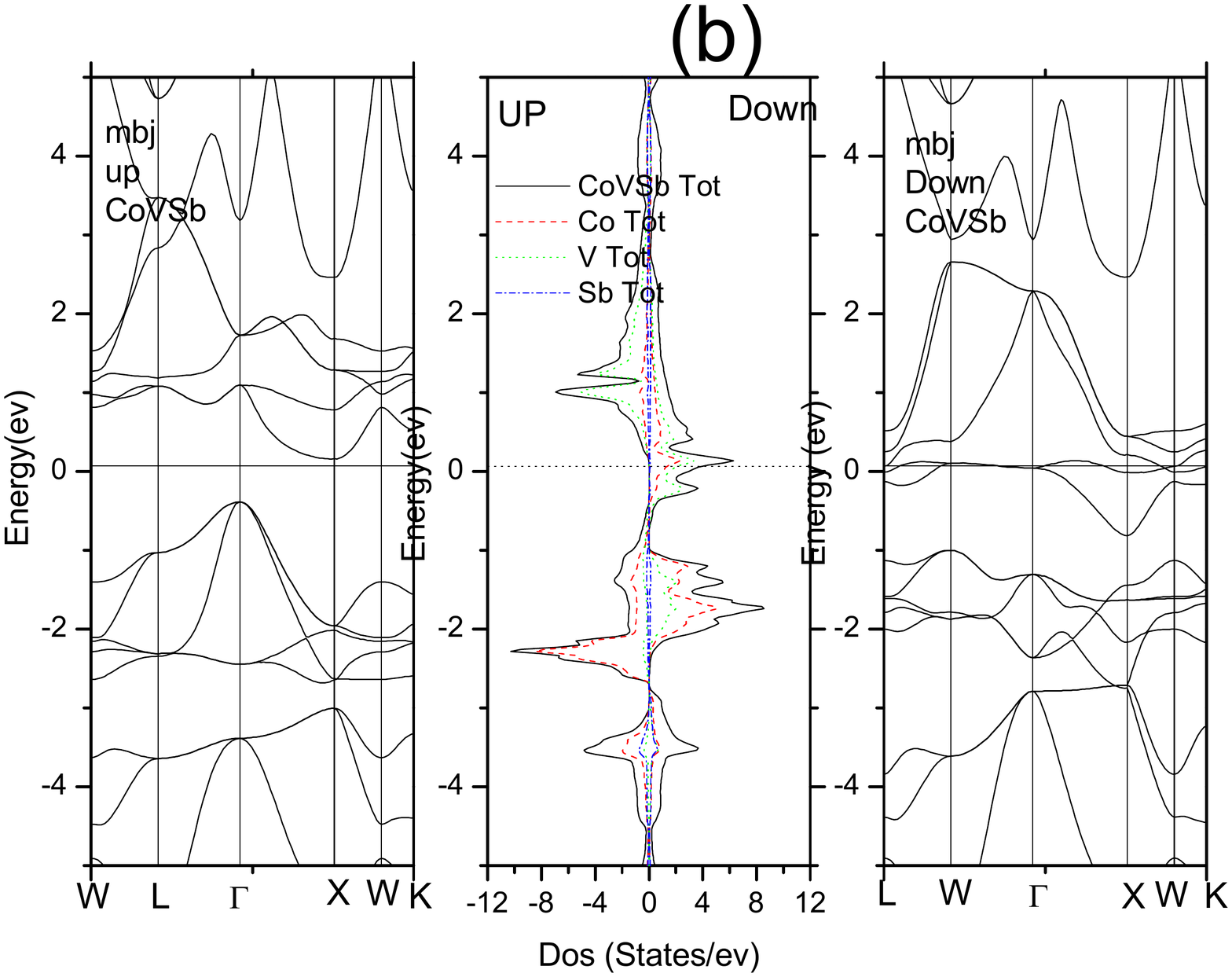}
\caption{(Colour online) The spin-projected total density of states of CoVSb with the corresponding band structures for (a) GGA and (b) TB-mBJ approximations.}
\label{fig4}
\end{figure}

\begin{figure}[!t]
\centering
\includegraphics[width=11.8 cm]{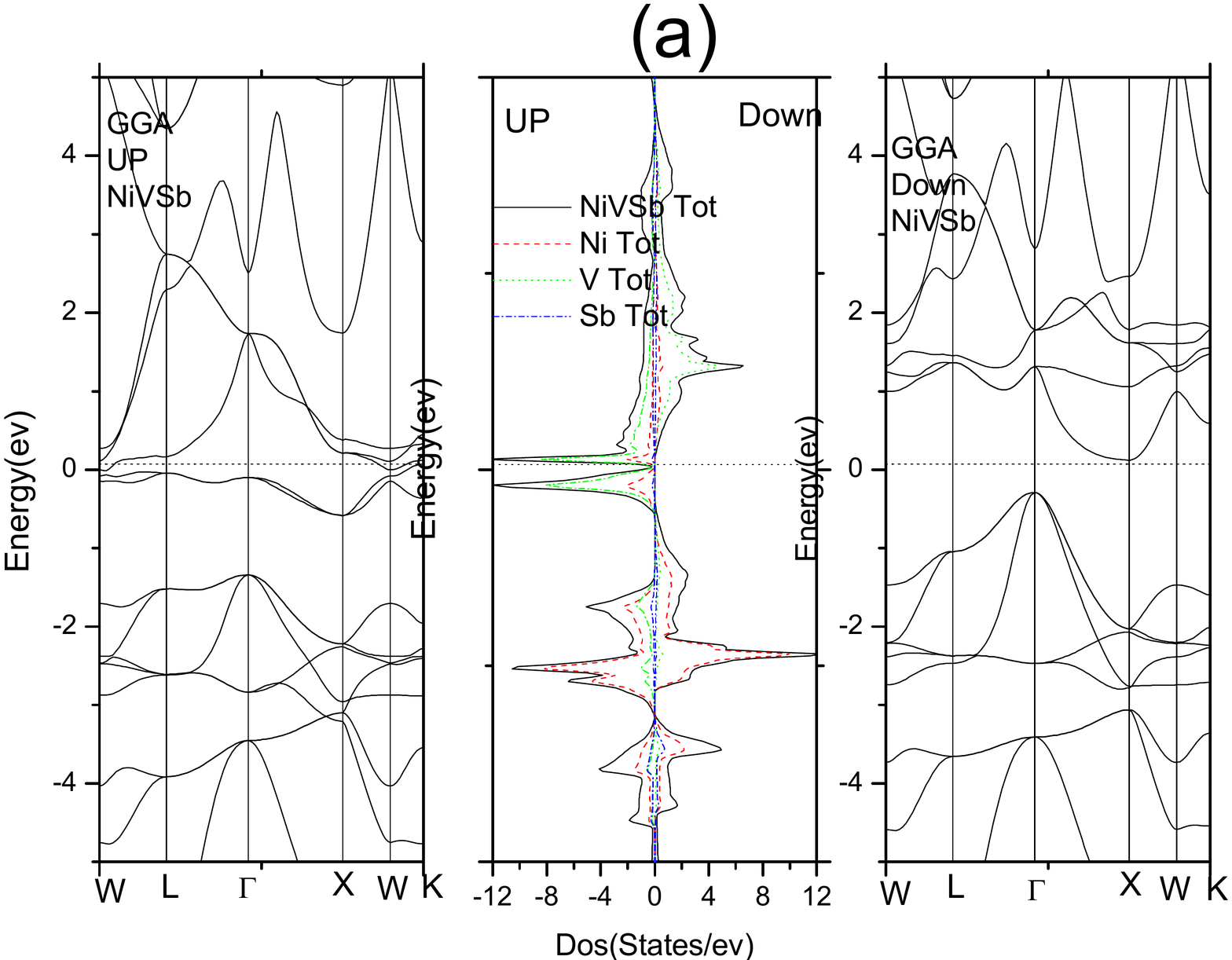}\\
\includegraphics[width=11.8 cm]{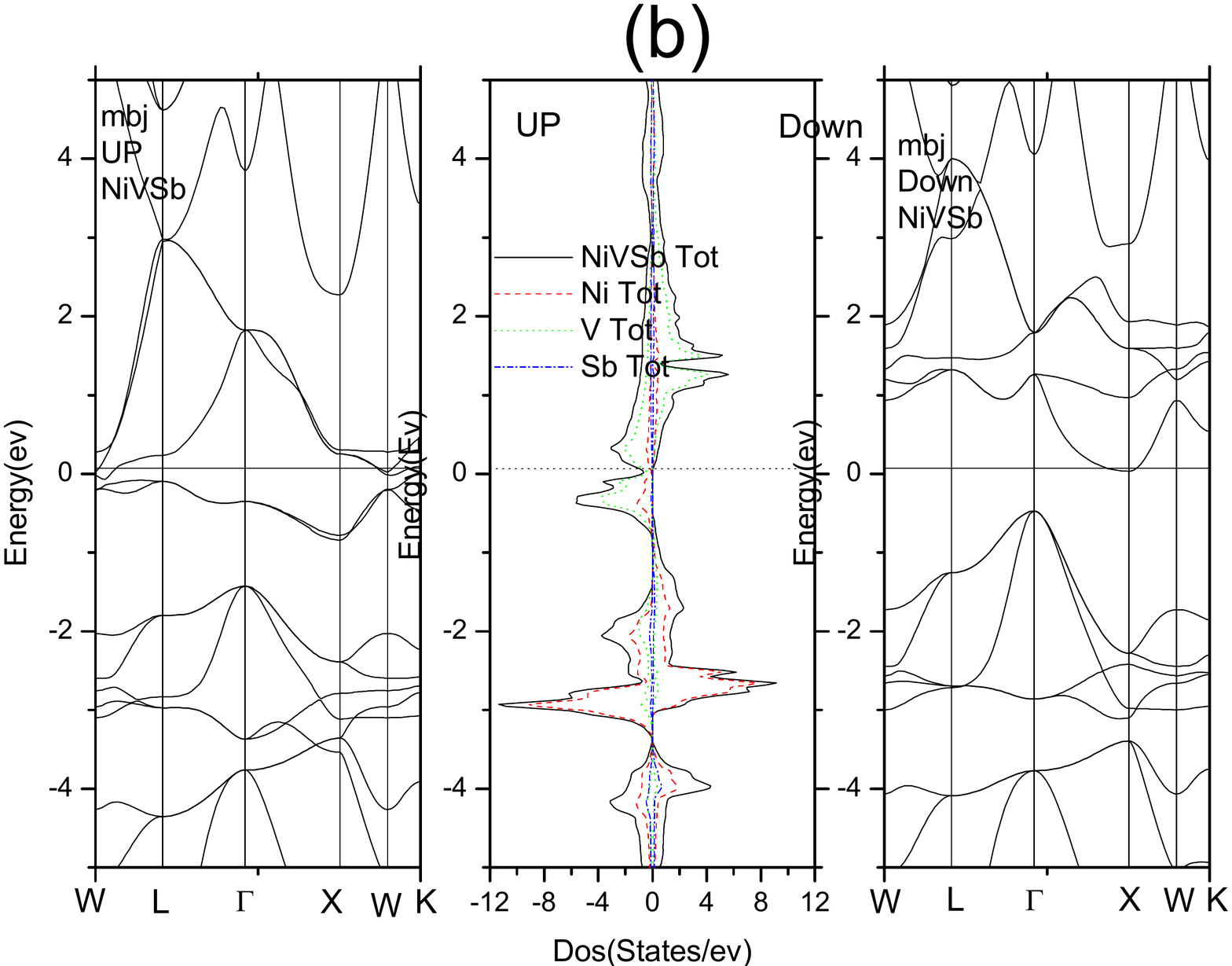}
\caption{(Colour online) The spin-projected total density of states of NiVSb with the corresponding band structures for (a) GGA and (b) TB-mBJ approximations.}
\label{fig5}
\end{figure}

The band structure and the density of states (DOS) of XVSb (X $=$ Fe, Co, Ni) half-Heusler compounds along the high symmetry directions in the first Brillouin zone are plotted in (figures~\ref{fig3}--\ref{fig5}), for GGA and TB-mBJ approximations at the equilibrium lattice parameter. The band gap values  with other theoretical data are presented in table~\ref{tab1}. 

The density of states of a system represents the number of states at every energy level. We have presented the total density of states (TDOS) and partial density of states (PDOS) of FeVSb in figures~\ref{fig3}~(a) and (b).  These figures show that there is symmetry between spin-up and spin-down  of the spin polarized DOS of FeVSb for both GGA and TB-mBJ approximations.
The compound presents an energy gap of $0.42$~eV at the Fermi level in both spin-up and spin-down band structure for GGA approximation and an energy gap of $0.74$~eV for TB-mBJ approximation which clearly explains that FeTiSb exhibits a semiconductor nature. The magnetic moments of Fe, V and Sb are $0.02047 \mu _\text{B}$, $-0.01486 \mu _\text{B}$ and $0.00010\mu _\text{B}$, respectively, for GGA calculations with an effective magnetic moment of $-0.00029\mu_\text{B}$. Meanwhile, the TB-mBJ approximation gives an effective magnetic moment of $0 \mu_\text{B}$.  

The calculated band structure for half-Heusler CoVSb compound is shown in figures~\ref{fig4}~(a) and (b). It can be seen for both GGA and TB-mBJ approximations that the majority-spin states show a semiconductor character, while the minority spin states indicate a metallic nature. The lowest conduction band at point~X and the highest valence band at the point $\Gamma$  form the band gap  and are equal to $0.80$~eV for GGA approximation and $0.92$~eV for mBJ.  	
Ab initio results propose that both the electronic and magnetic properties in these compounds are related to the apparition of the majority-spin gap. CoVSb alloy has 19~valence electrons and the total spin magnetic moment $M_\text{tot}$ is given by the relation $M_\text{tot}=(Z_\text{tot}-18)\mu_\text{B}$  for the half-Heusler alloys. Where $M_\text{tot}$ and $Z_\text{tot}$ are the total magnetic moment per formula unit and the number of total valence electrons, respectively. 
The spin magnetic moment is an integer and is equal to $1 \mu_\text{B}$ which confirms the HM behaviour for CoVSb with both GGA and mBJ approximations in good agreement with Slater-Pauling rule \cite{26, 27}. The individual magnetic moment of Co, Sb and V atoms are $0.72410\mu_\text{B}$,  $-0.04469\mu_\text{B}$ and $-1.54423\mu_\text{B}$ for mBJ approximation at normal pressure. The contribution of the atom~V to the magnetic moment of the compound  is greater compared to the other atom.

The band structures with the GGA and mBJ approximations of NiVSb  compound along the high symmetry directions in the Brillouin zone are presented in figures~\ref{fig5}~(a) and (b). It is clear that the maximum of the valence band is positioned at the point $\Gamma$ while the minimum of the conduction band is found at point X for both approximations. For this alloy, the (GGA) approximation predicts an indirect band gap of $0.36$~eV. However, for the mBJ potential approximation, NiVSb compound has an indirect band gap of $0.62$~eV at the $(\Gamma{-}\text{X})$ points.
To explain the band structure of our compounds, the density of states (DOS) of NiVSb compounds is calculated and plotted in figure~\ref{fig5} within the energy interval from $-5$ to $5$~eV. The obtained (DOS) shows a sharp peak at the
valence band and is located at $-2.38$~eV and $-2.91$~eV for GGa and mBJ approximations, respectively. These peaks  originate from Ni atoms. 
NiVSb alloy presents a spin magnetic moment of $2\mu_\text{B}$ per formula unit (see table~\ref{tab1}) for both GGA and mBJ approximations in good agreement with Slater-Pauling rule $(20-18=2)$.  For mBJ, Sb atom shows a negative magnetic moment of $-0.06005\mu_\text{B}$  ($-0.05196$ for GGA) per formula unit whereas Ni, Sb atoms give a positive magnetic moment of $0.11984\mu_\text{B}$ ($0.07698$ for GGA) and $1.73252\mu_\text{B}$ ($1.71898$ for GGA) per formula unit, respectively. 

\section{Conclusion}

The structural stability, electronic structures, magnetic properties,  and half-metallicity of half-Heusler XVSb (X $=$ Fe, Co and Ni) alloys have been investigated  by using a full-potential linearized augmented plane wave (FP-LAPW) method of density functional theory (DFT) within the modified Becke and Johnson potential (mBJ) exchange correlation potential and  the generalized gradient approximation (GGA).  
For different structures, the total energy calculations show that the most stable arrangement, where Fe, Co and Ni occupy the $(0,0,0)$ and V, Sb occupies the $(1/4,1/4,1/4)$ and $(3/4,3/4,3/4)$ positions, respectively.
The electronic band structure calculations show that the compounds CoVSb and NiVSb exhibit a half-metallic ferromagnetic (HMF) property with a magnetic moment of $1\mu_\text{B}$ and $2 \mu_\text{B}$ per formula unit at their equilibrium volume for both GGA and mBJ approximations. 
It is also found  that FeVSb presents a semiconductor nature with an energy gap of $0.41$~eV and $0.74$~eV for GGA and mBJ approximations, respectively. Moreover, negative value of the formation energy of these compounds confirms that these compounds are stable  and these compounds may be useful for spintronics applications.

\newpage

\ukrainianpart

\title{Теоретичне дослідження структурної стійкості, електронних і магнітних властивостей  XVSb ($\text{X}=\text{Fe}$, Ni і Co) напів-Гойслерівських сполук}
\author{M. Мохтарі\refaddr{label1,label2}, Ф. Даман\refaddr{label2},  Г. Бенабделла\refaddr{label3}, Л. Зекрі\refaddr{label1}, 
С. Беналья\refaddr{label2}, Н. Зекрі\refaddr{label1} }
\addresses{
\addr{label1} Унiверситет природничих наук i технологiй iм. Мохамеда Будiафа м. Оран, USTO-MB,  LEPM,\\ BP 1505, 31000 Оран, Алжир
\addr{label2} Інститут природничих наук і технологій, Університетський центр м. Тіссемсілт, 38000 Тіссемсілт, Алжир 
\addr{label3} Лабораторія обчислювальної фізики матерії, університет м. Сіді-Бель-Аббес 22000, Алжир
}

\makeukrtitle
\begin{abstract}
Структурні, електронні та магнітні властивості 	напів-Гойслерівських сполук
 XVSb (X $=$ Fe, Co і Ni) досліджено з використанням теорії функціоналу густини разом з узагальненим градієнтним наближенням (GGA) та Тран-Блаха модифікованим  Беке-Джонсон  (TB-mBJ) наближенням обмінного потенціалу. Встановлено, що  напівметалічні щілини  загалом достатньо розширені за рахунок  mBJ у порівнянні з наближенням GGA. Магнітні властивості  XVSb (X $=$ Fe, Co і Ni) добре визначені в межах  mBJ при точному цілому значенні магнітного моменту. Ширина заборонених зон, отримана з  TB-mBJ, добре узгоджується з наявними теоретичними даними.  FeVSb проявляє напівпровідниковий характер.  CoVSb і NiVSb показують  напівметалічну поведінку при сумарному магнітному моменті  $1\mu_\text{B}$ і $2\mu_\text{B}$, що добре узгоджується з правилом Слейтера-Паулінга.  Видається, що ці сплави є потенційними кандидатами для використання у спінтронних пристроях.
\keywords першопринципні розрахунки, напів-Гойслерівські сплави, структурні властивості, магнітні властивості
\end{abstract}

\end{document}